\documentclass{ws-procs975x65}

\usepackage{bm}

\begin{document}
%%%%%%%%%%%%%%%%%%%%%%%%%%%%%%%%%%%%%%%%%%%%%%%%%%%%%%%%%%%%%%%%%%%%%%%%%%%%%%%%%%%%%%%%%%%
\title{\uppercase{Inertial frames without the relativity principle: Breaking Lorentz symmetry}}
%%%%%%%%%%%%%%%%%%%%%%%%%%%%%%%%%%%%%%%%%%%%%%%%%%%%%%%%%%%%%%%%%%%%%%%%%%%%%%%%%%%%%%%%%%%
\author{\uppercase{Valentina Baccetti}, \uppercase{Kyle Tate}, and \uppercase{Matt Visser}}
%%%%%%%%%%%%%%%%%%%%%%%%%%%%%%%%%%%%%%%%%%%%%%%%%%%%%%%%%%%%%%%%%%%%%%%%%%%%
\address{School of Mathematics, Statistics, and Operations Research\\
Victoria University of Wellington \\
PO Box 600, Wellington 6140,
New Zealand}

\begin{abstract}
We investigate inertial frames in the absence of Lorentz invariance, reconsidering the usual group structure implied by the relativity principle.
We abandon the relativity principle, discarding the group structure for the transformations 
between inertial frames, while requiring these transformations to be at least linear (to preserve homogeneity). 
In theories with a preferred frame (aether), the set of transformations between inertial frames forms a 
groupoid/pseudogroup instead of a group, a characteristic essential to evading the von~Ignatowsky theorems. 
In order to understand the dynamics, we also demonstrate that the transformation rules for energy and momentum are in general affine. 
We finally focus on one specific and compelling model implementing a minimalist violation of Lorentz invariance. 
\end{abstract}

\bodymatter\bigskip  

Even though Lorentz symmetry seems to be very well established, the question of its possible breaking is still theoretically and 
experimentally engaging, and may have implications that span from quantum gravity to gravitational waves and cosmology.
There exist several different approaches to this topic in the literature, highlighting different aspects of the problem 
\cite{Mattingly:2005re,Colladay:1998fq,Anselmi:2011zz,Coleman:1998ti,Liberati:2001cr}.
In this work, we focus on a modification of von~Ignatowsky's 1910 argument that establishes  a tight relation between the 
relativity principle and the group structure for the Lorentz transformations \cite{Igna:1911b}. 
In particular, assuming linearity for the set of spacetime transformations between inertial frames $\{M_{i,j}\}$, isotropy of space time,
\emph{and the relativity principle}, the set $\{M_{i,j}\}$ will be either the Lorentz group for special relativity, or the Galilean group 
for classical Newtonian mechanics.

%%%%%%%%%%%%%%%%%%%%%%%%%%%%%%%%%%%%%%%%%%%%%%%%%%%%%%%%%%%%%%%%%%
%Kinematics
%%%%%%%%%%%%%%%%%%%%%%%%%%%%%%%%%%%%%%%%%%%%%%%%%%%%%%%%%%%%%%%%%%

Therefore, in order to explore the possibility of Lorentz symmetry breaking, we relax von~Ignatowsky's argument by 
abandoning the relativity principle, while retaining the concept of inertial frames. 
This implies that the set $\{M_{i,j}\}$ is no longer a group and that we are {\em de facto} introducing a preferred frame, the {\em aether}.
Since we want to preserve as much as possible of the definition of inertial frames, we want the transformations $\{M_{i,j}\}$ to map straight lines
into straight lines, a condition that is satisfied if the transformations are linear. We therefore  define the transformation from the 
aether frame $F$ to any other inertial frame $\bar{F}$ as:
\begin{equation}
\label{trans}
 M = \left[\begin{array}{c|c} \gamma & -\gamma \bm{u}^T \\ \hline 
-\Sigma \bm{v}&  \Sigma \end{array} \right]; \qquad
\left[\begin{array}{c} \bar t \\ \bm{\bar x} \end{array}\right] = M \; \left[\begin{array}{c}  t \\ \bm{x} \end{array}\right].
\end{equation}
Here $M$ is in general a function of $\bm{v}$ but it can also depend on the orientation of $\bar{F}$ with respect to $F$. 

From this, it can easily be deduced that the velocity $\bm{v}_{\mathrm{aether}}$  of the aether as seen from the inertial frame $\bar{F}$, 
and the velocity  $\bm{v}_{\mathrm{moving}}$ of the frame $\bar{F}$ as seen from the aether, are not {\em equal-but-opposite} and may not even be collinear.

Indeed:
\begin{align}
\bm{v}_{\mathrm{aether}}=-\frac{\Sigma \; \bm{v}_{\mathrm{moving}}}{\gamma}.
\end{align}
Even though the set of transformations $\{M_{F,\bar{F}}\}$ no longer forms a group, a rather interesting result is that the set 
of transformations between two inertial frames $F_1$ and $F_2$, $\{M(F_2,F_1)\}=\{M(F_2)M^{-1}(F_1)\}$ 
presents a groupoid/pseudogroup structure. That is, this set it is closed under \emph{partial product}, has an identity and an inverse, it is associative.
It is exactly the technical difference between a group and groupoid that is essential to side-step von~Ignatowsky's argument.

%%%%%%%%%%%%%%%%%%%%%%%%%%%%%%%%%%%%%%%%%%%%%%%%%%%%%%%%%%%%%%%%%%
%Dynamics
%%%%%%%%%%%%%%%%%%%%%%%%%%%%%%%%%%%%%%%%%%%%%%%%%%%%%%%%%%%%%%%%%%

In order to explore how energy and momentum transform under these new conditions, 
we need to introduce some notion of dynamics, that is, we need to be able to define Newton's three laws in each inertial frame.
In this respect, the most appropriate tool is the Lagrangian/Hamiltonian formalism, which is also essential for the 
formulation of a possible quantum theory.
We therefore set up a Lagrangian $L(\bm{\dot{x}})$, or a Hamiltonian $H(\bm{p})$, for each inertial frame. $L(\bm{\dot{x}})$ and $H(\bm{p})$ are only 
defined up to some constant terms since the Euler--Lagrange equations are not affected by boundary terms.  

To obtain the transformation laws for energy and momentum we impose the condition that the extrema of an action, defined in one 
inertial frame, must coincide to the extrema of the action calculated in any other inertial frame, at least up to boundary terms. In view of the 
groupoid structure of $\{M_{i,j}\}$, there is no loss of generality in considering a moving frame $\bar{F}$
and the aether frame $F$ for which we can write (directly in the Hamiltonian formalism):
\begin{equation}
  \int \left\{(\bar{E}+\bar{\epsilon})-(\bm{\bar{p}}+\bm{\bar{\pi}})\cdot\left({d\bm{\bar{x}}\over d\bar{t}}\right)\right\}d\bar{t}
  =
  \int \left\{(E+\epsilon)-(\bm{p}+\bm{\pi})\cdot\left({d\bm{x}\over dt}\right)\right\}dt.
\end{equation}
The quantities $(\epsilon, -\bm{\pi}^T)$ and $(\bar{\epsilon}, -\bm{\bar{\pi}}^T)$ represent the intrinsic ambiguities due to possible boundary terms. 
This yields the following transformation rule for energy and momentum (for details see reference \refcite{Baccetti:2011aa})
\begin{equation}
 (\bar{E}+\bar{\epsilon},-\bm{\bar{p}^T}-\bm{\bar{\pi}^T})=(E+\epsilon,-\bm{p}^T-\bm{\pi}^T)\; M^{-1}.
\end{equation}
These are {\em affine} transformations, with the affine piece depending only on the intrinsic ambiguities. 
If we define the four component vector $P=(E,-\bm{p}^T)$, we can see that it transforms in the 
dual space of the four-component vector $X=(t,\bm{x}^T)^T$.

Legitimate questions arise concerning the status of the  the invariant quantities usually defined in special relativity. In order
to disentangle this matter we consider a particle moving in the aether with velocity $\bm{v}$ and with energy $E(\bm{p})$. 
If we now go to the particle rest frame $\bar{F}$, with $\bar{\bm{v}}=0$, we can associate a rest-energy
$\bar{E}=E_0(\bar{F})$ and a rest-momentum $\bm{\bar{p}}=\bm{p}_0(\bar{F})$, that will depend on the state of motion of $\bar{F}$ 
with respect to the aether. The rest-momentum $\bm{p}_0(\bar{F})$ may in general not be equal to zero, 
as for example explained in reference~\refcite{Baccetti:2011us}. 

Transforming back to the aether frame (where, at least to a first approximation, we carry out our experiments), and setting the offset terms, without any loss of generality, as $\epsilon=\bar{\epsilon}=\pi=0$ and $\bar{\pi}=-\bm{p}_0$, we can construct the following expression:

\begin{equation}
\label{inva}
 E^2-\varpi ||\bm{p}||^2c^2=\gamma^2(1-\varpi c^2 ||\bm{u}||^2)E^2_0(\bar{F}).
\end{equation}
Here we have introduced a constant $c$ with the dimension of a velocity (not necessarily the speed of light) and some arbitrary
function $\varpi(\bar{F})$.

From this we can see that the concept of rest energy still exists but it now depends on the state of motion of $\bar{F}$
with respect to the aether, through $E_0(\bar{F})$. This result is completely different from the special relativity case, where the rest-energy and the 
rest-momentum are intrinsic properties of the particle that cannot depend on its velocity with respect to the aether.
Such a counterintuitive outcome is due to the fact that the functional form for $E(\bm{p})$, and for the transformations $M(\bm{v})$ between inertial
frames,  are in general independent of each other.

This property is what allows us to explore a possible {\em minimalist Lorentz-violating model}, where we consider the spacetime to 
be Lorentz invariant, i.e. the physics of clocks and rulers is Lorentz invariant, while we allow for a Lorentz-violation in the 
energy-momentum sector, i.e. for one sub-sector of the particle spectrum (e.g. neutrinos). 
In this case we see that the quantity \eqref{inva} simplifies to 
\begin{equation}
\label{inva2}
 E^2-||\bm{p}||^2=E^2_0({\bar{F}}).
\end{equation}
This quantity is now invariant when transforming from one inertial frame to another, but still depends on the absolute state of 
motion of $\bar{F}$ with respect to the aether:
\begin{equation}
 \bar{\bar{E^2}}-||\bar{\bar{\bm{p}}}||^2=E^2-||\bm{p}||^2=E^2_0({\bar{F}}).
\end{equation}
In summary: Even in the absence of a relativity principle, and even in the absence of Lorentz symmetry, there is still quite a lot that can usefully be said about the transformation laws between inertial frames,  and their implications for both kinematics and dynamics.

%%%%%%%%%%%%%%%%%%%%%%%%%%%%%%%%%%%%%%%%%%%%%%%%%%%%%%%%%%%%%%%%%%%%%%
\section*{Acknowledgments}
%%%%%%%%%%%%%%%%%%%%%%%%%%%%%%%%%%%%%%%%%%%%%%%%%%%%%%%%%%%%%%%%%%%%%%
VB acknowledges support by a Victoria University PhD scholarship. \\
MV acknowledges support by the Marsden Fund, and by a James Cook fellowship, both administered by the Royal Society of New Zealand. 

%%%%%%%%%%%%%%%%%%%%%%%%%%%%%%%%%%%%%%%%%%%%%%%%%%%%%%%%%%%%%%%%%%%%%%

%%%%%%%%%%%%%%%%%%%%%%%%%%%%%%%%%%%%%%%%%%%%%%%%%%%%%%%%%%%%%%%%%%%%%%

\begin{thebibliography}{10}
%%%%%%%%%%%%%%%%%%%%%%%%%%%%%%%%%%%%%%%%%%%%%%%%%%%%%%%%%%%%%%%%%%%%%%

\bibitem{Mattingly:2005re}
  D.~Mattingly,
  %``Modern tests of Lorentz invariance,''
  Living Reviews in Relativity {\bf 8} (2005) 5
  %[gr-qc/0502097].
  %%CITATION = GR-QC/0502097;%%
  %300 citations counted in INSPIRE as of 21 Feb 2013
  
\bibitem{Colladay:1998fq}
  D.~Colladay and V.~A.~Kostelecky,
  %``Lorentz violating extension of the standard model,''
  Phys.\ Rev.\ D {\bf 58} (1998) 116002
  %[hep-ph/9809521].
  %%CITATION = HEP-PH/9809521;%%
  %979 citations counted in INSPIRE as of 21 Feb 2013
  
\bibitem{Anselmi:2011zz}
  D.~Anselmi,
  %``Renormalization and Lorentz symmetry violation,''
  PoS CLAQG {\bf 08} (2011) 010.
  %%CITATION = POSCI,CLAQG08,010;%%
  
\bibitem{Coleman:1998ti}
  S.~R.~Coleman and S.~L.~Glashow,
  %``High-energy tests of Lorentz invariance,''
  Phys.\ Rev.\ D {\bf 59} (1999) 116008
  %[hep-ph/9812418].
  %%CITATION = HEP-PH/9812418;%%
  %841 citations counted in INSPIRE as of 21 Feb 2013
  
\bibitem{Liberati:2001cr}
  S.~Liberati, T.~A.~Jacobson, and D.~Mattingly,
  ``High-energy constraints on Lorentz symmetry violations'',
   Proceedings of the 2nd meeting on CPT and Lorentz symmetry, edited by V.~Alan Kostelecky. (World Scientific, Singapore, 2005). 
  %hep-ph/0110094.
  %%CITATION = HEP-PH/0110094;%%
  

\bibitem{Igna:1911b}
    W.~von~Ignatowsky,
    %"Das Relativitätsprinzip". 
    Archiv der Mathematik und Physik {\bf 17} (1911) 1--24
 
\bibitem{Baccetti:2011aa}
  V.~Baccetti, K.~Tate, and M.~Visser,
  %``Inertial frames without the relativity principle,''
  JHEP {\bf 1205} (2012) 119
  %[arXiv:1112.1466 [gr-qc]].
  %%CITATION = ARXIV:1112.1466;%%
 
  
\bibitem{Baccetti:2011us}
  V.~Baccetti, K.~Tate, and M.~Visser,
  %``Lorentz violating kinematics: Threshold theorems,''
  JHEP {\bf 1203} (2012) 087
  %[arXiv:1111.6340 [hep-ph]].
  %%CITATION = ARXIV:1111.6340;%%
  
  
  
%%%%%%%%%%%%%%%%%%%%%%%%%%%%%%%%%%%%%%%%%%%%%%%%%%%%%%%%%%%%%%%%%%%%%%
\end{thebibliography}
\end{document}